\documentclass[prb,twocolumn,showpacs]{revtex4} 
\usepackage{graphicx}

\def\prru{PrRu$_4$P$_{12}$}
\def\prfe{PrFe$_4$P$_{12}$}
\def\ga{\Gamma}
\input epsf
\epsfverbosetrue
\tighten
\begin{document}
\draft
%\twocolumn[\hsize\textwidth\columnwidth\hsize\csname @twocolumnfalse\endcsname

\title{Theory of the Metal-Insulator Transition in PrRu$_4$P$_{12}$ and
\prfe}
 
\author{S. H. Curnoe$^{1,2}$\cite{Stephanie},
H. Harima$^{3}$, K. Takegahara$^4$ and K. Ueda$^{1,5}$}
\address{$^{1}$Institute for Solid State Physics, University of Tokyo, Kashiwanoha 5-1-5, Kashiwa,
Chiba 277-8581, Japan\\
$^2$Department of Physics and Physical Oceanography, Memorial University of
Newfoundland, St. John's NL, A1B 3X7, Canada\\
$^3$Institute of Scientific and Industrial Research, Osaka
University, Ibaraki, Osaka 567-0047, Japan\\
$^4$Department of Materials Science and Technology, Hirosaki
University, Hirosaki, Aomori 036-8561, Japan\\
$^5$Advanced Science Research Center, Japan Atomic Energy Research
Institute, Tokai, Ibaraki 319-1195, Japan}
 
\date{\today}
 
%\maketitle
 
\begin{abstract}
All symmetry allowed couplings between the $4f^2$-electron
ground state doublet of trivalent
praseodymium in
\prru\ and \prfe\ and displacements of the phosphorus, iron or ruthenium ions
 are considered.
Two types of displacements can
change the crystal lattice from
body-centred cubic to simple orthorhombic or 
to simple cubic. The first type  lowers the point
group symmetry from tetrahedral to orthorhombic, while the second
type leaves it unchanged,
with corresponding space group reductions 
Im$\bar3 \rightarrow $Pmmm and Im$\bar3 \rightarrow $Pm$\bar3$ respectively.
In former case, the lower
point-group symmetry splits the degeneracy of the
$4f^2$ doublet into states with opposite quadrupole moment, which then leads
to anti-quadrupolar ordering, as in \prfe.  Either kind of displacement
may conspire with nesting
of the Fermi surface
to cause the metal-insulator or partial metal-insulator
transition observed in \prfe and \prru.   We
investigate this scenario using band-structure calculations, 
and it is found that displacements of the phosphorus ions
in \prru
(with space group reduction Im$\bar3\rightarrow$Pm$\bar3$)
open a gap everywhere on the
Fermi surface.

\end{abstract}
 
%]

\maketitle

\narrowtext
%\tightenlines

\section{Introduction}
The rare earth-filled skutterudites (RT$_{4}$Pn$_{12}$; R=rare earth;
T = Fe, Ru or Os; Pn = P, As or Sb) exhibit a wide range of phenomena,
including superconductivity in
nearly all of the La-compounds\cite{meis1981,shir1997,take2000}, as well as
%LaFe$_4$P$_{12}$\cite{meis1981},
%LaRu$_4$P$_{12}$\cite{meis1981}, LaRu$_4$As$_{12}$\cite{shir1997},
%LaRu$_4$Sb$_{12}$\cite{shir1997,take2000},
%LaOs$_4$P$_{12}$,
PrRu$_4$As$_{12}$\cite{shir1997} and
PrRu$_4$Sb$_{12}$\cite{take2000}, heavy fermion superconductivity in
PrOs$_4$Sb$_{12}$\cite{baue2002}, semiconducting or semimetallic behaviour in the Ce-compounds\cite{gran1984}
and
ferromagnetism in several of the Nd- and Eu-compounds\cite{meis1981,take2000,gran1984,seki2000}
%NdFe$_4$P$_{12}$, NdRu$_4$P$_{12}$\cite{meis1981},
%NdRu$_4$Sb$_{12}$\cite{take2000}(should check this)
%EuFe$_4$P$_{12}$\cite{gran1984}, EuRu$_4$P$_{12}$\cite{seki2000}
%EuRu$_4$Sb$_{12}$\cite{take2000}
and UFe$_4$P$_{12}.$\cite{meis1985}
PrRu$_4$P$_{12}$ undergoes a metal-insulator (M-I) transition at
approximately
$T_{MI}=60$K,\cite{seki1997,nanb1999} while in PrFe$_4$P$_{12}$  there is a partial M-I transition
at a much lower temperature, seen
as a sharp upturn in the resistivity at $T_{MI} = 6.7$K,\cite{tori1987,sato2000b} which
then tends to zero at low temperatures. In
both cases, the M-I transition is accompanied by a structural phase transition
which doubles the volume of the primitive cell.\cite{lee2001,iwas2001}

%PrFe$_4$P$_{12}$ exhibits Kondo-like behaviour, with  a minimum in the
%resistivity at 200K.\cite{sato2000b}  
%A meta-magnetic transition is
%observed in the magnetisation\cite{tori1987} and specific
%heat.\cite{mats2000b,aoki2002} 
%At large magnetic fields the
%specific heat coefficient $\gamma$ is very large, greater than 1 J/K$^2\cdot$mol,\cite{mats2000b} and
%the cyclotron masses are also strongly enhanced.\cite{suga2001}
In 
\prfe, a peak in the magnetic
susceptibility is observed at $T_{MI}$,\cite{meis1981}
and $T_{MI}$ is
field-dependent, but there is no magnetic ordering.\cite{kell2001}
However, in-field neutron
diffraction experiments found direct evidence of anti-quadrupolar
ordering below $T_{MI}$.\cite{iwasa}
In contrast, $T_{MI}$ of
PrRu$_4$P$_{12}$ is not field dependent (as seen in
the specific heat\cite{seki2000b} and thermal expansion\cite{matsu2000})
and no peak is observed
in the susceptibility either,\cite{seki1997}
which suggests that there is no ordering of any kind at $T_{MI}$.
An upturn in the susceptibility at very low temperatures
hints that there may be some kind of ordering
below $T=0.35$K,\cite{meis1981} and is likely related to the
low-temperature magnetic field dependent peaks found in the thermal
expansion and specific heat, whose positions
 decrease in temperature with decreasing
field.\cite{matsu2000,seki2000b}

Nesting of the Fermi surface may be the common feature of the M-I
transition in both of these 
materials.  Band-structure calculations 
for
LaFe$_4$P$_{12}$,\cite{suga2000} PrFe$_4$P$_{12}$ and PrRu$_4$P$_{12}$\cite{hari2002}
found an  approximately cubed-shaped hole-like Fermi surface,
with nesting wavevector $q=(1,0,0)$ (due to the
48th band in LaFe$_4$P$_{12}$ and the 49th band in PrFe$_4$P$_{12}$
and PrRu$_4$P$_{12}$). 
The band itself has a roughly flat dispersion, which means that slight
differences can cause
substantially different Fermi surface topologies.
Thus,
in PrRu$_4$Sb$_{12}$ and LaRu$_4$Sb$_{12}$ the resulting FS does not have the
nesting property.\cite{hari2002,mats2002}
LaFe$_4$P$_{12}$ and PrFe$_4$P$_{12}$
possess an additional smaller hole-like spherical Fermi surface,
so that the M-I transition in
PrFe$_4$P$_{12}$ is incomplete, and it remains a metal at low temperatures. 
It has been suggested that
a nested Fermi surface is also a possible cause of the resistivity upturns
seen in NdFe$_4$P$_{12}$,
SmRu$_4$P$_{12}$, GdRu$_4$P$_{12}$ and TbRu$_4$P$_{12}$.\cite{matsu2002b}

If nesting is a prerequisite for the M-I transition, then so is
a cell-doubling structural instability,
which halves the Brillouin
zone (BZ) at the nesting wave vector.
All of the rare earth-filled skutterudites crystallise in the
body-centred cubic (bcc) lattice with space group Im$\bar 3$ (\#204), except for
the low-temperature phases of PrFe$_4$P$_{12}$ and PrRu$_4$P$_{12}$.
In these cases, superlattice reflections are a clear indication of doubling of the
primitive cell.\cite{lee2001,iwas2001}
The change in structure appears to be due to a displacement of Fe ions
in PrFe$_4$P$_{12}$ and both of Ru and
P ions in PrRu$_4$P$_{12}$, with space groups
Pmmm (\#47) and Pm$\bar 3$ (\#200) respectively. 

In Section II,
all symmetry allowed
couplings of Pr$^{3+}$ $4f^2$ electrons to the lattice are considered. 
It is shown that 
anti-quadrupolar ordering of the $4f^2$ electrons follows when 
the point group symmetry at the Pr site is lowered from
tetrahedral to orthorhombic (as is the case in the
Im$\bar 3 \rightarrow$ Pmmm transition).
In Section III we present
band structure calculations which show that a gap opens
everywhere at the  Fermi energy when P ion displacements (with space
group reduction Im$\bar 3$ $\rightarrow$ Pm$\bar3$) are
considered in PrRu$_4$P$_{12}$.\cite{curnoe}
We summarise our results in Section IV.

\section{Structural Phase Transition and Anti-Quadrupolar Ordering}

According to XANES studies on \prru\cite{lee1999} and susceptibility measurements
on PrFe$_4$P$_{12}$,\cite{tori1987} the valency of the Pr ion is $+3$, ie, a $4f^2$ configuration
with total angular momentum $J=4$.
The Ru/Fe ions form a crystal field with O$_h$ symmetry with
respect to the Pr ions, but the P ions
lower the symmetry to T$_h$.
Under O$_h$ symmetry, the $J=4$ state splits into a singlet,
a doublet and two triplets,  corresponding to
the representations $\ga_1 \oplus \ga_3 \oplus \ga_4 \oplus\ga_5$ of O respectively.
Under the actual symmetry  T$_h$, the $\ga_3$ doublet is split into
states related by time reversal symmetry
(complex conjugates).
Crystal
field splitting alone cannot determine the ground state, but
specific heat measurements on \prfe\cite{aoki2002} and
\prru\cite{seki2000b}
favour the doublet,
which is
\begin{eqnarray}
|\Gamma_3^+\rangle &=& \sqrt{7/24}|4\rangle -\sqrt{5/12}|0\rangle
+\sqrt{7/24}|-4\rangle \\
|\Gamma_3^-\rangle &= & \sqrt{1/2}|2\rangle + \sqrt{1/2}|-2\rangle.
\end{eqnarray}
These states carry quadrupole moment 
$\langle \ga_3^{\pm}|3J_z^2 - J^2|\ga_3^{\pm}\rangle = \pm 8\hbar^2$.

On the other hand, a singlet ground state, 
\begin{equation}
|\ga_1\rangle = \sqrt{5/24}|4\rangle -\sqrt{7/12}|0\rangle
+\sqrt{5/24}|-4\rangle
\end{equation}
with a low lying triplet first excited state has not been conclusively
ruled out.  The exact form 
of the triplets is unknown in $T_h$, because in general they are
linear combinations of
\begin{eqnarray}
|\ga_4^x\rangle & = & [|3\rangle+|-3\rangle+\sqrt{7}(|1\rangle+|-1\rangle)]/4 \\
|\ga_4^y\rangle & = & i[-|3\rangle+|-3\rangle+\sqrt{7}(|1\rangle-|-1\rangle)]/4\\
|\ga_4^z\rangle & = & (|4\rangle - |-4\rangle)/\sqrt{2}
\end{eqnarray}
and
\begin{eqnarray}
|\ga_5^{yz}\rangle & = & [\sqrt{7}(-|3\rangle +|-3\rangle)-|1\rangle+|-1\rangle]/4 \\
|\ga_5^{xz}\rangle & = & i[-\sqrt{7}(|3\rangle +|-3\rangle)+|1\rangle +|-1\rangle]/4\\
|\ga_5^{xy}\rangle & = & (|2\rangle - |-2\rangle)/\sqrt{2}.
\end{eqnarray}

In the following
discussion, we often refer to the representations of O$_h$, but since
the actual crystal symmetry is T$_h$, we should use the following
correspondences between the representations of O and T:
%\begin{table}[ht]
\begin{center}
\begin{tabular}{|c|c|c|} \hline
Dimension & O & T \\ \hline
1&$\ga_1$ & $\ga_a$ \\
1&  $\ga_2$  & $\ga_a $ \\
2&  $\ga_3$   & $\ga_b\oplus \ga_c$  \\
3&  $\ga_4$   &$\ga_d $ \\
3&  $\ga_5$   &$\ga_d $ \\ \hline
\end{tabular}
\end{center}
%\caption{Corrspondance between the representations O and T.  The number in
%the first column indicates the dimension.}
%\end{table}
We use the representations of O$_h$ for convenience, especially in the
discussion of the displacement modes of the Fe/Ru ions, since by themselves
they form a crystal field with O$_h$ symmetry.   However, 
the above table should always be used to reduce the
representations of O$_h$ to those of T$_h$. 

Now we consider all possible modes of the displacements.
There are eight Fe/Ru ions
within the 
bcc conventional cell, which are found half-way between
neighbouring Pr ions
(the centres of the hexagonal faces of
the bcc  Wigner-Seitz (primative) cell).  
They form corner-sharing cubes about each Pr ion.
There are 24 displacement modes, 
which transform according to the
representations $\ga_1\oplus\ga_2'\oplus\ga_3\oplus\ga_3'\oplus\ga_4\oplus 2\ga_4'
\oplus 2\ga_5\oplus \ga_5'$ of O$_h$, where the primes indicate the
odd ({\em ungerade}) representations. 

Twelve
P ions are found {\em inside} the bcc  primitive cell (note that the bcc
primitive cell is half the volume of the bcc conventional cell).  
The 36 modes of the P ion
displacements transform according
to the representations $2(\ga_a\oplus \ga_b\oplus \ga_c)\oplus 4\ga_d \oplus \ga_a'\oplus
\ga_b' \oplus \ga_c'\oplus 5\ga_d'$ of T$_h$.
Each of the 36 modes on a single site can be either in-phase or
anti-phase with the modes on the  nearest neighbours, for 
a total of 72 modes in the bcc conventional cell. 
However, only the anti-phase modes will cause a
cell-doubling; the in-phase modes may lower the point group symmetry but
the lattice structure will remain body-centred.

Next, we consider all symmetry allowed couplings between the $4f^2$ angular
momentum
and the lattice.
The angular momentum cannot couple to the odd representations, and since
$f^{\dagger}f$ transforms as
$\ga_3\otimes\ga_3 = \ga_1\oplus \ga_2 \oplus \ga_3$
we find that only the
Fe/Ru lattice distortions which can couple to the $4f^2$ $\ga_3$ doublet
are the $\ga_1$ and $\ga_3$ modes.
The $\ga_1$ mode
is a dilatation of the cube with the eight Fe/Ru
ions on each corner and
is represented by the displacement vector $v = (1,1,1)$.  $v$ is shorthand 
notation for the 24-component vector of the positions for each of the eight
ions.    The $\ga_1$ mode
does not lower the point group symmetry, but
it does double the primitive
cell and change the lattice from bcc to simple cubic (sc). 
The two-dimensional $\ga_3$ mode is represented by
the vectors $v=(1,1,-2)$
and $v=(1,-1,0)$.  Again, this is shorthand notation for the distortion
of the cube with eight Fe/Ru ions on the corners into a solid rectangle.
Such a distortion also changes the
lattice, and lowers the point group symmetry from O$_h$ to
D$_{4h}$ or D$_{2h}$ respectively.  Under O$_h$ crystal field symmetry,
there is only one coupling for each of the $\ga_1$ and $\ga_3$ modes, since
the identity ($\ga_1$) only appears once in the decompositions
$\ga_1\otimes \ga_3\otimes\ga_3 = \ga_1\oplus \ga_2\oplus\ga_3$ and
$\ga_3\otimes \ga_3 \otimes \ga_3 = \ga_1\oplus\ga_2\oplus3\ga_3$.  Under the
actual symmetry T$_h$,
$\ga_2$ reduces to $\ga_a$, and there
are two couplings for each mode instead of one.

There are more choices of coupling to the P ion displacements.
There are two different $\ga_a$ modes, represented by $v=(1,1,1)$, 
and each of these has two couplings to the
$4f^2$ doublet.  Also there are two kinds of $\ga_b\oplus \ga_c$ modes, 
represented by $v=(1,1,-2)$ and $v=(1,-1,0)$,
which altogether have four different
couplings  to the $4f^2$ doublet.

Although any of the above couplings might lead to a M-I transition via
coupling to the conduction electrons,
there is a  significantly different consequence for those distortions
which lower the
point group symmetry (namely the $\ga_3$ or $\ga_b\oplus\ga_c$ 
modes) compared to  those which do not (the $\ga_1$ or $\ga_a$ modes).
When the point group symmetry is lowered by the $\ga_3$ or $\ga_b\oplus\ga_c$
modes,
the degeneracy of the $4f^2$ doublet is lifted,
and as we shall show next, the result is
anti-quadrupolar ordering.

We consider first the $\ga_3$ or $\ga_a\oplus \ga_b$ modes.
The general form of the Hamiltonian which describes
the coupling between Fe/Ru ion or P ion displacements and the $4f^2$ doublet
in the  bcc conventional cell is
\begin{eqnarray}
H &=& \epsilon(f^{\dagger}f
+ f^{*\dagger}f^{*}+f^{'\dagger}f'+ f^{*\dagger}f^{*}) \nonumber \\
& & +\frac{1}{2}[T(f^{\dagger}f^{*}A^*-f^{'\dagger}f^{'*}A^*) + \mbox{h.c.}]
+ \frac{\omega}{2} |A|^2 \nonumber \\
& & + \frac{B}{4}|A|^4+ \frac{C_1}{6}|A|^6+\frac{C_2}{12}(A^6+A^{*6}) \nonumber \\
&  & +
\frac{C_3}{12}(A^6-A^{*6}).  \label{H} 
\end{eqnarray}
The operators $f^{\dagger}$ and $f'^{\dagger}$ 
act on the two neighbouring Pr sites and correspond 
to the complex wavefunctions
$|f\rangle = (|\Gamma_3^+\rangle  -
i |\Gamma_3^-\rangle)/\sqrt{2}$ and $|f^{*}\rangle = (|\Gamma_3^+\rangle  +
i |\Gamma_3^-\rangle)/\sqrt{2}$.
$A$ is the complex 
amplitude of the displacement $\vec{v}=\vec{v}_3+\chi \vec{v}_2+\chi^2 \vec{v}_1$
of either eight Fe/Ru ions or 24 P ions,
where $\chi = \exp\left(\frac{i2\pi}{3}\right)$ and the
displacement vectors are $\vec{v}_3=(-1,-1,2)$, $\vec{v}_2=(-1,2,-1)$ and
$\vec{v}_1=(2,-1,-1)$.  
The actual displacement is then 2Re$A\vec{v}$.
$A$ corresponds to the
anti-phase mode as described above, therefore it changes sign under the
bcc lattice translation $t=(\frac{a}{2},\frac{a}{2},\frac{a}{2})$, hence
odd order in $A$ terms do not appear in $H$.  (Note that under the same translation
$f\rightarrow f'$).
All of the coupling constants are real, except for
$T$, which is real in O$_h$, but complex
in T$_h$
(therefore, there are twice as many coupling constants under T$_h$,
as noted above).  
The $C_3$  term is
prohibited in O$_h$, but allowed in T$_h$.
$C_1$ should be greater than
$C_2$ and $C_3$ for stability.

We begin by diagonalising $H$ with respect to $f$.
Writing
$A= a e^{i\alpha}$ and $T=t e^{i\tau}$, 
the eigenvalues are $\epsilon \pm ta$ with
eigenvectors $e^{i(-\alpha+\tau)}|f\rangle \pm  |f^{*}\rangle$.
Then the ground state projection of $H$ is
\begin{equation}
H_{eff} = -ta 
+\frac{\omega}{2} a^2 + \frac{B}{4}a^4+
\frac{C_1}{6}a^6+\frac{C_2}{6} a^6\cos 6\alpha + \frac{C_3}{6}a^6\sin 6\alpha  .
\end{equation}
We will assume for the moment that 
$\tau=0$ and $C_3=0$ (as in O$_h$ symmetry).
Then, for  $C_2<0$, 
$H_{eff}$ has minima at
%$a = \frac{-B+\sqrt{B^2 +4(t-\omega)(C_1-C_2)}}{2(C_1-C-2)}$
$\alpha = 0, \pi/3, 2\pi/3, \pi, 4\pi/3, 5\pi/3$.  The corresponding displacements
are proportional to 
$\pm(-1,-1,2)$, $\pm(-1,2,-1)$ and $\pm(2,-1,-1)$. 
%Note that this result does not
%depend on the choice of basis $v_i$ given above.
These displacements reduce the crystal field symmetry from
O$_{h}$ to D$_{4h}$, thereby splitting  the
$4f^2$ doublet,  so that now the ground state is non-degenerate.
For example, 
for the $(-1,-1,2)$
displacement ($\alpha = 0$),
the $4f^2$ ground state is
 $|\Gamma_3^{-}\rangle$. 
The $4f^2$ ground state on the neighbouring site 
is orthogonal to this (because of the change of sign
on $T$),
therefore the $4f^2$ ground state alternates between $|\Gamma_3^{+}\rangle$ and
$|\Gamma_3^{-}\rangle$ from site to site.
Since the states
$|\Gamma_3^{\pm}\rangle$ carry opposite quadrupole
moments, it follows that
the ground state will have anti-quadrupolar ordering. 
%of the form $3J_z^2-J^2$.
The displacement doubles the primitive cell and
changes the structure from bcc to simple orthorhombic
(Im${\bar 3}$ to Pmmm).

When $C_2>0$, $H_{eff}$ has  minima at $\alpha = \pi/6, \pi/2, 5\pi/6,
7\pi/6, 3\pi/2, 11\pi/6$.   This leads to domains with 
displacements of the form $\pm(1,-1,0)$, $\pm(1,0,-1)$ and $\pm(0,1,-1)$
The corresponding $4f^2$ ground states carry quadrupole moments $J_x^2-J_y^2$ 
{\em etc.} which alternate sign from site to site in the same way as
for $C_2<0$.    The space
group reduction is also the same: Im${\bar 3}$ to Pmmm.

When the coupling constant
$T$ is complex and $C_3$ is non-zero (as they are in general in T$_h$ symmetry),
the angle $\tau$ is non-zero and the angle $\alpha$ is no longer
fixed.
The ground state
can be a mixture of $|\ga_3^{+}\rangle$ and $|\ga_3^{-}\rangle$ and
the displacements are a mixture of $(2,1,1)$ and $(1,-1,0)$ modes.
However, for the case when the $4f^2$ doublet is more strongly coupled
to the Fe/Ru displacements,
the O$_h$ symmetry is broken only by the presence of the P ions, which
interact indirectly through their contribution to the crystal field.
Therefore, in this case, it may be appropriate to take the
O$_h\rightarrow$ T$_h$ symmetry lowering as a perturbative effect, and
then the resulting anti-quadrupole moment should be close to the maximum
value which is
found under O$_h$ symmetry.

Now we consider the $\ga_1$ or $\ga_a$ modes.  The Hamiltonian
is
\begin{eqnarray}
H &=& \epsilon(f^{\dagger}f
+f'^{\dagger}f'+ f^{*\dagger}f^{*} + f'^{*\dagger}f'^{*}) \nonumber \\
& & + TA(f^{\dagger}f -f'^{\dagger}f' + \mbox{h.c.})
+ \omega A^{2} \nonumber \\
& & + \frac{B}{4}A^4 +\frac{C}{6}A^6, 
\end{eqnarray}
where $A$ is now the operator for the displacement $v=(1,1,1)$ and
all of the coupling constants are real.
In this case, 
the  $4f^2$ ground state remains degenerate 
and there is no lattice distortion.   Therefore, in order for this type of
distortion to occur, the source of the instability
must be something else, such as 
coupling to conduction electrons, as we show in Section III.

The assumption that the doublet is the ground state of the Pr$^{3+}$ ion
is not a strict requirement.  In the alternate scenario
of a singlet ground state and low-lying triplet excited state,
Kiss and Fazekas\cite{fazekas} showed how to 
couple the $\ga_3$ lattice distortion mode to the quadrupole moment of
the triplet.  In this case, the Hamiltonian takes the same form as in 
(\ref{H}) except that the operator $f$ corresponds to the wavefunction
$|f\rangle = (|\ga_4^z\rangle+\chi|\ga_4^y\rangle+\chi^2
|\ga_4^x\rangle)/\sqrt{3}$.  Diagonalising $H$ breaks the degeneracy of the triplet; 
the energy of one state
is lowered by $TA$, and another is raised by the same amount. If $TA$ is 
large enough  then the state $e^{i(-\alpha+\tau)}|f\rangle - |f^*\rangle$
is the ground state of the system.  
Note that this state carries no magnetic moment, but in general has a 
quadrupole 
moment.  Anti-quadrupolar ordering follows as before.

Thus, beginning with all possible couplings between the Pr$^{3+}$ $4f^2$
doublet and the lattice, we find that there are only two types of
lattice distortions that are consistent with a cell-doubling
structural phase transition. 
In the first case ($\ga_3$), the point group symmetry is lowered from
T$_h$ to D$_{2h}$, and the doublet is split into states carrying
opposite quadrupole moment.
Anti-quadrupolar ordering is a consequence of the lower point group symmetry and
doubling of
the primitive cell. 
For the second kind of cell-doubling
distortion ($\ga_1$),
the point group symmetry remains T$_h$, and
there is no quadrupole ordering.
Experimental evidence
suggests that the first scenario applies to PrFe$_4$P$_{12}$ and while the 
second one applies to PrRu$_4$P$_{12}$. 

%On the other hand,  an upturn of the
%susceptibility\cite{meis1981} and magnetic field dependent
%peaks of the specific heat \cite{seki2000b}  and thermal expansion
%coefficient \cite{matsu2000} similar
%to \prfe near $T_{MI}$
%are found at temperatures much lower than $T_{MI}$
%in PrRu$_4$P$_{12}$. 

\section{Band structure calculations}

Band structure calculations were performed  using the FLAPW-LDA+U
method\cite{hari2001} to investigate the origin of the M-I transition in 
PrRu$_4$P$_{12}$.   The LDA+U method has described the non-magnetic localized $f$ electron system.\cite{hari2002}

%\subsection{\prru}
Various $\ga_1$-type lattice
distortions involving P ion displacements were considered.
The space group is lowered from Im$\bar{3}$ to Pm$\bar{3}$ as the primitive
cell is doubled.
In Im$\bar{3}$,
we used the observed lattice parameters, i.e. the lattice constant $a=8.0424$\AA\ and the internal parameters  $u=0.3576$ and $v=0.1444$ for P positions.\cite{sekineprivate}
In Pm$\bar{3}$,
the positions of Ru ions are equivalent as $8i$, $(1/2+\delta_T,1/2+\delta_T,1/2+\delta_T)$.
While the P ions occupy the inequivalent  positions $12j$, $(0,u+\delta^A_u,v+\delta^A_v)$ and $12k$, $(1/2,1/2+u+\delta^B_u,1/2+v+\delta^B_v)$. 
When $\delta_T=0$, $\delta^A_u=\delta^B_u$ and $\delta^A_v = \delta^B_v$,
the space group has the higher symmetry Im$\bar{3}$ with the smaller 
primitive cell.
The site of Pr ions is split to two sites ($1a$ (0,0,0) and $1b$ (1/2,1/2,1/2)) in Pm$\bar{3}$ due to the inequivalent surroundings.
However, since the local symmetry m$\bar{3}$ for the both sites is unchanged,
any kind of lifting the degeneracy of  the localized $4f^2$ state is not expected in Pm$\bar{3}$.
Here the singlet $\Gamma_1$ states for $4f^2$ electrons as the starting state were assumed and  found after the self-consistent steps.

Although Ru ions are experimentally distorted to minimize the total energy,
the position of Ru ions does not affect the Fermi surface,
because the Fermi surface of PrRu$_4$P$_{12}$ consists of mainly P-$p$ band.
Therefore $\delta_T =0$ is assumed in the calculations.
Moreover, the anti-phase distortion of P ions are considered.
Several values of the two parameters $\delta_u = \delta^A_u = -\delta^B_u$ and $\delta_v = \delta^A_v = -\delta^B_v$ with $\delta_u \delta_v<0$ were considered in the calculations,
and are listed  in Table \ref{tableI}.

\begin{table}

\caption{The calculated band gaps $\Delta_E$ in eV with distortions.
The minus sign means that the valence and conduction bands are overlapped.
The asterix indicates that a self-consistent solution was not obtained.}
\label{tableprp}
\begin{center}
\begin{tabular}{rrrrrrrr} \hline
$\delta_v \backslash  \delta_u$ & 0.000 & 0.001 & 0.002 & 0.003 & 0.004 & 0.005 & 0.006 \\ \hline
-0.010&  0.4& --& --& --& --& --& -\\
-0.005& -3.2& -3.1&  0.8& 1.8& 2.0& 1.9&-- \\
-0.004& -4.4& -2.0& -0.3& 1.1& 1.6& 1.6&-- \\
-0.003& -5.5& -3,2& -1.5& -0.1& 1.0& 1.2&-- \\
-0.002& -6.6& -4.4& -2.6& -1.2& -0.1& 0.6&-- \\

-0.001& -7.8& -6.3& -3.8& -2.3& -1.2& *&-- \\
 0.000& --     & -6.9& -4.9& -3.5& -2.4& -1.7&-1.2 \\
 0.001& -7.8& -7.5& -6.1& -4.3& -3.6& -2.8&-- \\ \hline
\label{tableI}
\end{tabular}
\end{center}
\end{table}

\begin{figure}
%\vspace{.2in}
\begin{center}
\includegraphics[width=3.4in]{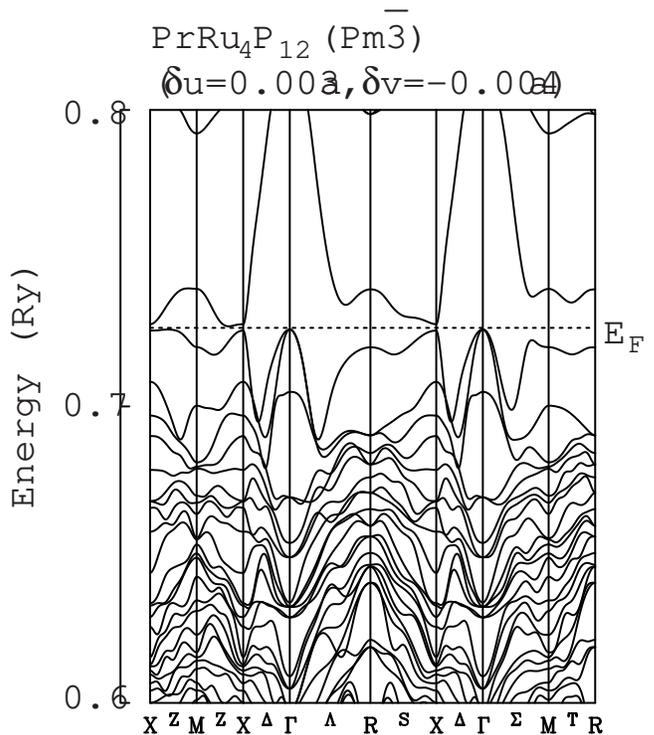}
%\includegraphics{34prru4p12.eps}
%\vspace{.2in}
\caption{\label{prp}Insulating band structure for PrRu$_4$P$_{12}$ obtained with $\delta_u=0.003a$ and $\delta_v=-0.004a$.}
\end{center}
\end{figure}

\begin{figure}
%\vspace{.2in}
\begin{center}
\includegraphics[width=3.4in]{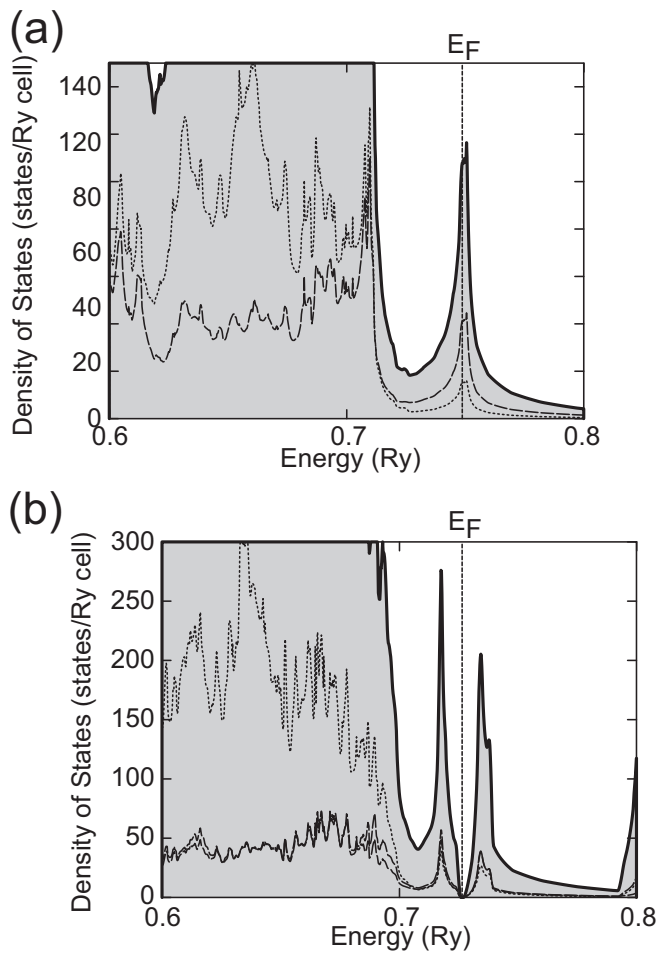}
%\includegraphics{34prru4p12.eps}
%\vspace{.2in}
\caption{\label{dos}The density of states  for PrRu$_4$P$_{12}$ for (a) Im${\bar 3}$  and (b) Pm${\bar 3}$  with $\delta_u=0.003a$ and $\delta_v=-0.004a$.
Solid line, dotted line and dashed line indicates total, Ru-$d$ component and P-$p$ component, respectively.
Note that there are two P-$p$ components in (b).}
\end{center}
\end{figure}

Optimal results were obtained for $\delta_u=0.003a$ and $\delta_v=-0.004a$. 
This distortion produced a gap across the entire Fermi surface, as shown in Fig.~\ref{prp}.
Fig.~\ref{dos} shows the density of states in the vicinity of the Fermi level both for Pm${\bar 3}$ and Im${\bar 3}$  corresponding to Fig~\ref{prp}.

%\begin{widetext}
%\vspace{.2in}
%\vspace{.2in}
%\end{widetext}

%\begin{widetext}
%\vspace{.2in}
%\vspace{.2in}
%\end{widetext}

\section{Discussion and Summary}
There are significant differences in the nature of the
transition between \prfe\ and PrRu$_4$P$_{12}$.
The M-I transition temperatures differ by an order of magnitude,
with $T_{MI}=6.7$K for PrFe$_4$P$_{12}$
\cite{sato2000b} versus $T_{MI}=60$K for 
PrRu$_4$P$_{12}$ \cite{seki1997}.  In addition, $T_{MI}$ of the Fe-compound
decreases in temperature with increasing magnetic
fields \cite{aoki2002,mats2000b}, while the Ru-compound shows {\em no}
magnetic field dependence of $T_{MI}$ (seen as a jump in the
specific heat)
\cite{seki2000b}, and no anomaly in the magnetic susceptibility at
$T_{MI}$ \cite{seki1997} either. 
This suggests that, in spite of the similarities in
the shape of the Fermi surface and in the structural phase
transition,  the lattice instability
may be of a quite different character between the two materials.

The observation of anti-quadrupolar ordering below $T_{MI}$ in \prfe\ is strong
evidence that the structural phase transition in \prfe\ involves
lattice distortion modes which lower
the point group symmetry from T$_h$ to D$_{2h}$ (the $\ga_3$ modes). 
Moreover, the experiments indicate that it is the Fe ions which are
displaced in this mode.\cite{iwas2001}
As for PrRu$_4$P$_{12}$, the absence of magnetic anomalies
mentioned at the beginning of this section,
as well as Raman studies\cite{sekine1999} and electron diffraction\cite{lee2001},
indicate that the structural phase transition at $T_{MI}$ does not
change the local Pr site symmetry and 
is not accompanied by quadrupole ordering, hence
it involves those modes which do not lower the point group symmetry
(the $\ga_1$ modes).   Our band structure calculations
favour P ion displacements over Ru ion displacements
since the former are shown to be
a source of the M-I transition.

The low temperature upturn in the susceptibility and the magnetic
field dependence of the thermal expansion and specific heat in
\prru\ at low temperatures suggest 
that there may be an ordering
at a lower temperature.
It could be due to a
second structural phase transition which lowers the point group and
lifts the degeneracy of the doublet, but
so far there are no experimental clues about the
type of deformation involved.   There are two possibilities,
either in-phase or anti-phase
modes of the $\ga_3$ type distortions described in detail above. 
In-phase modes were excluded as a source of the M-I transition because
they do not double the primitive cell.  Since the cell is already doubled below
the first transition, exclusion of the in-phase modes is no longer
necessary.  Then the space group is lowered
through the sequence Im$\bar3\rightarrow$Pm$\bar3\rightarrow$Pmmm,
but now the $4f^2$ state has quadrupolar ordering below the
second transition. 
Only the P-ions participate in the in-phase modes, but the anti-phase modes
may involve either P ions or  Fe/Ru ions.
The anti-phase modes lead
to anti-quadrupolar ordering and lower the space group through
the same sequence Im$\bar3\rightarrow$Pm$\bar3\rightarrow$Pmmm.
Given the similarities between the transition temperatures and
the magnetic field
dependences of structural phase transition in \prfe\
and the proposed second transition in PrRu$_4$P$_{12}$,
it seems more likely that the
second transition in \prru\ would be
 due to displacements of the Ru ions, rather than
either in-phase or anti-phase modes of P ions.

To summarise, we have studied structural phase transitions coupled to localised
$4f^2$ states and to band electrons in PrFe$_4$P$_{12}$ and
PrRu$_4$P$_{12}$.   In both cases the source of the structural
instability is identified, but the mechanisms are quite different.  In
PrFe$_4$P$_{12}$,  a Jahn-Teller type mechanism occurs,
as coupling between $\ga_3$-type  displacements and
the $4f^2$ electrons lifts the degeneracy and 
lowers the energy of the  ground state of the $4f^2$ electrons.  The
space group is reduced from Im$\bar{3}$ to Pmmm.  Coupling
to the band electrons removes part of the Fermi surface.
In \prru,  our bandstructure calculations show that coupling 
between lattice displacements and band electrons will remove
the Fermi surface everywhere and produce a metal-insulator transition. 
The space group is lowered from Im$\bar{3}$ to Pm$\bar{3}$.   This produces no
change in symmetry at the Pr sites, hence no orbital ordering
occurs.

%Other routes to anti-quadrupolar ordering are possible.
%SmRu$_4$P$_{12}$ also undergoes a partial M-I transition at $T_{MI} =16$K,
%accompanied by anti-ferromagnetic (AF) ordering,\cite{matsu2002b}
%which is due to splitting of the
%crystal field ground state quartet
%of Sm$^{3+}$ $4f^5$ electrons into doublets which
%carry quadrupole moments.\cite{curnoe2} A nested Fermi surface which
%vanishes under a $(1,1,-2)$-type lattice distortion is
%also expected ed for this material.
%The transition appears as a sharp upturn in the resistivity at
%$T_{MI}$, which then decreases to
%a minimum at 12K, similar to the behaviour seen in PrFe$_4$P$_{12}$.  Below
%10K  the resistivity increases rapidly.  A slight hump is seen
%just below $T_{MI}$, which grows into a peak under magnetic field,
%while the jump at $T_{MI}$ shrinks,
%and similar behaviour is observed in the specific heat.
%A minimum in the resistivity at 50K is attributed to the Kondo effect.

%A different crystal field level scheme has been proposed for the origin of
%anti-quadrupolar ordering in PrFe$_4$P$_{12}$.\cite{fazekas}  Mixing
%between a $\ga_1$
%ground state with a $\ga_4$ first excited state can also give rise to
%anti-quadrupolar ordering.

%\begin{figure}[ht]
%\epsfysize=2.9in
%\epsfbox[60 240 570 670]{test2.ps}
%\caption{
%}
%\end{figure}

\acknowledgements
S. C. was supported by the Japan Society for the Promotion of Science.  We
thank Y. Aoki, K. Matsuhira, C. Sekine and I. Sergienko
for valuable discussions.

\end{document}